\documentclass[letterpaper]{article}
\usepackage[T1]{fontenc}
\usepackage{amsmath}
\usepackage{subcaption}
\usepackage{multirow}
 \usepackage{geometry}
\usepackage{setspace}
\usepackage[style=chem-acs, articletitle=true]{biblatex}
\usepackage{graphicx}
\usepackage{float}
\newfloat{scheme}{htbp}{los}
\floatname{scheme}{Scheme}
\floatname{chart}{Chart}
\newfloat{graph}{htbp}{loh}
\usepackage{chemformula} 
\usepackage[version = 4]{mhchem} 
\AtEveryBibitem{%
  \clearfield{note}%
}
\AtEveryBibitem{%
  \clearfield{pages}%
}
\AtEveryBibitem{%
  \clearfield{volume}%
}
\usepackage{authblk}
\author[1]{Baptiste Martin}
\author[2]{Shukai Yao}
\author[1]{Chunyu Li}
\author[3]{Anthony Bocahut}
\author[4]{Matthew Jackson}
\author[1]{Alejandro Strachan$^{*}$}
\affil[1]{School of Materials Engineering and Birck Nanotechnology Center, 
Purdue University, West Lafayette, Indiana 47907, United States}
\affil[2]{Department of Chemical and Biomolecular Engineering, 
Vanderbilt University, Nashville, TN 37235, United States}
\affil[3]{Syensqo Speciality Polymers, Sain-Fons, 69190, France}
\affil[4]{Syensqo Composite Materials, Piedmont, South Carolina 29673, United States}

\title{Predictive models for strain energy in condensed phase reactions}
\date{*strachan@purdue.edu}

\begin{document}

\maketitle







\begin{abstract}
Molecular modeling of thermally activated chemistry in condensed phases is essential to understand polymerization, depolymerization, and other processing steps of molecular materials. Current methods typically combine molecular dynamics (MD) simulations to describe short-time relaxation with a stochastic description of predetermined chemical reactions. Possible reactions are often selected on the basis of geometric criteria, such as a capture distance between reactive atoms. Although these simulations have provided valuable insight, the approximations used to determine possible reactions often lead to significant molecular strain and unrealistic structures. We show that the local molecular environment surrounding the reactive site plays a crucial role in determining the resulting molecular strain energy and, in turn, the associated reaction rates. We develop a graph neural network capable of predicting the strain energy associated with a cyclization reaction from the pre-reaction, local, molecular environment surrounding the reactive site. The model is trained on a large dataset of condensed-phase reactions during the activation of polyacrylonitrile (PAN) obtained from MD simulations and can be used to adjust relative reaction rates in condensed systems and advance our understanding of thermally activated chemical processes in complex materials.
\end{abstract}


\section{Introduction}
Molecular modeling of reactive processes in condensed phases is critical for developing predictive models for various technologically important processes, from the curing of thermosets \cite{schichtel_modeling_2020, varshney_molecular_2008-1, wu_atomistic_2006} and the polymerization of thermoplastics, \cite{ma_understanding_2023, wang_coarse-grained_2010, lempesis_atomistic_2017, laurien_atomistic_2018, lee_coarse-grained_2009,hossain_molecular_2010}, to the processing of molecular materials such as carbon fibers \cite{desai_molecular_2017,yao_molecular_2024, abbott_atomistic_2011, saha_carbonization_2012, shi_high-temperature_2021,  shi_generation_2021, yao_ordered_2023} and degradation processes \cite{wang_influence_2018, sayko_degradation_2020, karuth_reactive_2022}. The resulting molecular structures can be used with molecular dynamics (MD) simulations to predict thermal \cite{shenogina_molecular_2012,sellan_size_2010,bandyopadhyay_molecular_2011} mechanical \cite{li_molecular_2011,li_free_2016, fitzer_optimization_1986,odegard_modeling_2005, han_molecular_2007}, and transport properties \cite{zhang_polymer_2014}. These efforts and many other recent publications highlight the growing importance of modeling processing steps and generating accurate atomic structures for molecular materials. In this paper, we introduce a graph neural network (GNN) to improve predictions of the associated reaction rates, which are central to many of these models.

Although MD simulations with reactive force fields, such as ReaxFF \cite{nielson_development_2005}, can, in principle, describe these processes, the limited simulation time (typically nanoseconds) severely limits their applicability. Despite this limitation, several studies explored processing reactions using reactive MD. For example, Vashish et al. \cite{vashisth_accelerated_2018} simulated the crosslinking between an amine and an epoxy. To observe these rare events within the achievable time scales, they employed a temperature more than 100 K higher than in experiments and only reached low conversion degrees. Similarly, Saha and Schatz investigated the carbonization of Polyacrylonitrile(PAN) using ReaxFF \cite{saha_carbonization_2012}. Their simulations were conducted at 2500 K, a temperature significantly exceeding the experimental range of 1300 K to 2000 K \cite{yang_preparation_2019, hao_highly_2018, cipriani_crosslinking_2016}. Although such elevated temperatures can accelerate reactions, they also create conditions different from experiments, potentially altering relative reaction rates between competing mechanisms and reaction pathways. An alternative approach builds on the separation of timescales between molecular relaxation processes and chemical reactions. For example, models combine non-reactive MD simulations\cite{yao_molecular_2024, rennekamp_hybrid_2020,gissinger_modeling_2017} to relax the molecular system with a stochastic description of discrete chemical events. These methods use geometrical criteria to identify possible reactions in condensed phase systems out of a predetermined set of possibilities. Reactions are selected from a list, and the topology describing the covalent interactions is updated to reflect the selected chemical reactions. MD simulations are used between chemical reaction steps to relax and thermalize the system. The main limitation of these methods is that the rates of possible chemical reactions are described with simple rules that do not account for the local molecular environment that can hinder or facilitate the reaction process. This can result in unrealistically high strain energies and structures, especially at high conversion degrees. 

\noindent With the ultimate objective of developing improved models for the reaction rates in condensed molecular systems, this paper introduces a machine learning method capable of predicting the local strain-energy-associated cyclization reactions in PAN from the pre-reaction configuration. We use a graph to describe the local molecular structure and found that a message-passing neural network is capable of accurate predictions, enough to disincentivize  most of the reactions that would result in high local strains.

\section{Methods} 
\label{Methods}
\subsection{Data acquisition}
\begin{figure}[H]
\centering
\includegraphics[width=1\textwidth]{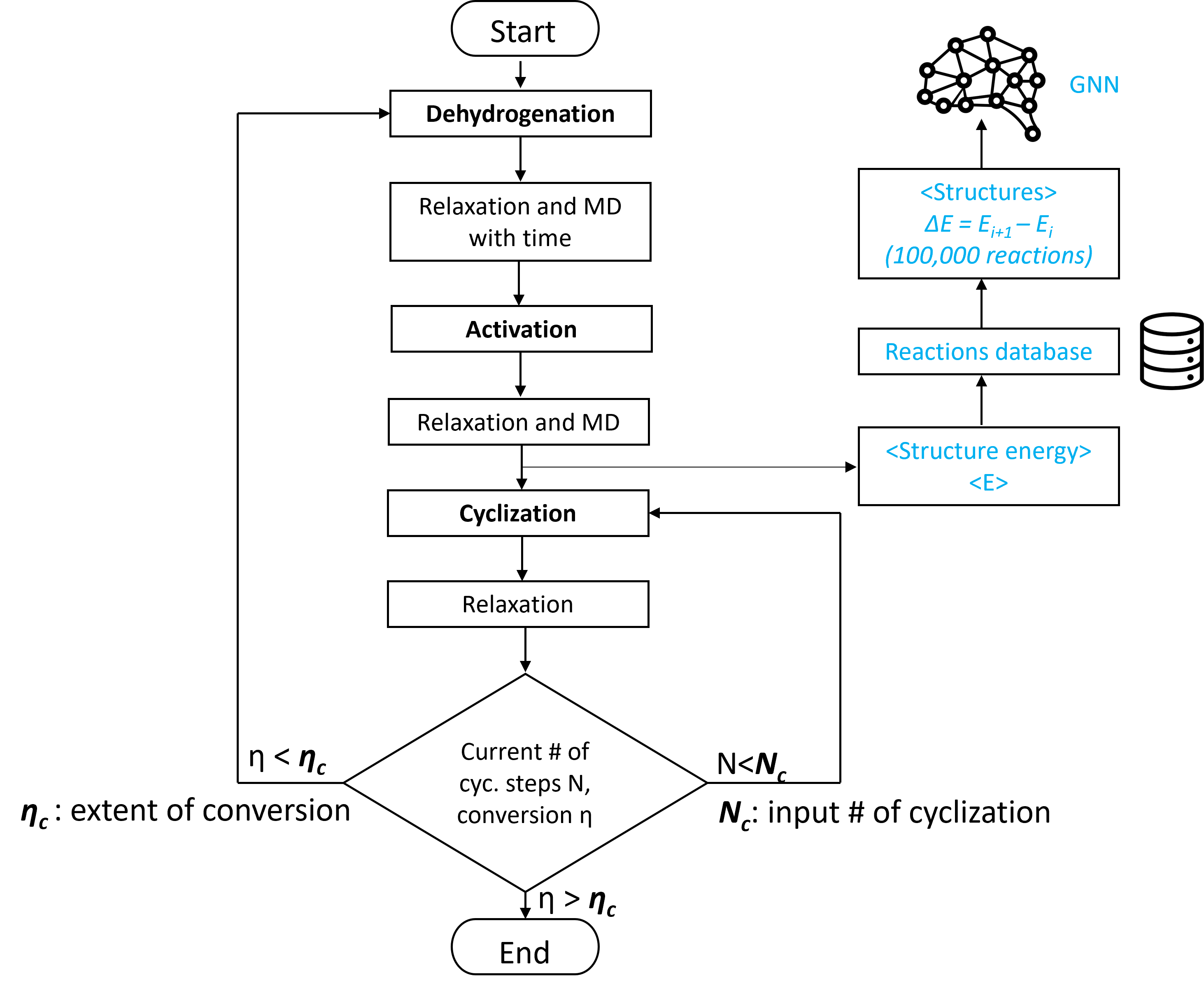}
  \caption{Flowchart illustrating the PAN cyclization steps (black), as introduced in \cite{yao_molecular_2024}, and the data extraction process used to train the GNN (blue). This procedure generates a database of all reactions, linking molecular structures and chemical reactions to their corresponding energies.
  }
  \label{fig1:Shukai}
\end{figure}

Data used to train our GNN models (local strain energies and molecular structures) were obtained from prior simulations of the stabilization of PAN \cite{yao_molecular_2024}. As illustrated in fig.\ref{fig1:Shukai} the process is represented sequentially, although dehydrogenation and activation occur randomly at different carbon sites. Konstantopoulos et al., \cite{konstantopoulos_introduction_2020} reported that dehydrogenation occurs first then followed by cyclization, the latter being initiated through a free radical mechanism as shown in \cite{yao_molecular_2024}. In this work, we use the term ‘activation’, to denote the radical formation that triggers cyclization. For the sake of simplicity, the overall mechanism is therefore considered sequential.
\newpage
\noindent The model integrates stochastic chemistry steps, governed by geometric criteria with molecular mechanics and MD for relaxation between successive reactions. Simulations are initialized with 40 chains of PAN, composed of 600 monomers each and proceed until conversion, defined as the fraction of nitrile groups that have reacted with a carbon atom to form new bonds or cyclized PAN units, reaches 90\%, as shown in Fig. \ref{fig1:Shukai}.
To obtain local strain energy and molecular geometries of the reactive sites (fig.\ref{fig:SystemA}, \ref{fig:System B}) for the different models, we extracted the atomic coordinates of each reaction site and calculated the per-atom energies of the involved atoms during 10 ps MD simulations, both before and after the reaction. The total energy per atom includes contributions from bond stretching, angles, torsions, improper torsions, and pairwise non-bonded interactions. Expressing the energy storage on a per-atom basis facilitates straightforward comparisons between models with different numbers of atoms. For each reactive site, we then calculated the difference between the energies before and after cyclization. This difference, referred to as $\Delta E$ or strain energy in this paper, it represents the local strain associated with the reaction.
To train the GNN models, the data are stored using the extended XYZ file format, which contains the atomic coordinates before the reaction, the box size, forces set to zero, and the strain energy.
It should be emphasized that the only post-reaction information included in the dataset is the strain energy, defined as the energy difference between the post- and pre-reaction configurations
see Eq.~\eqref{eq:strain_energy}.
\begin{equation}
    \text{Strain energy} = \Delta E  = E_{post} - E_{pre}
    \label{eq:strain_energy}
\end{equation}
Only relative energy differences are physically meaningful, since the Hamiltonian of the system changes upon reaction due to modifications in covalent bonding. As a result, the absolute energies of pre- and post-reaction states cannot be directly compared. The strain energy $\Delta E$ therefore serves as the relevant physical descriptor for training the GNN models. 
\subsection{Models}
\begin{figure}[H]
    \begin{subfigure}[b]{\textwidth}
        \centering
        \includegraphics[scale=0.5]{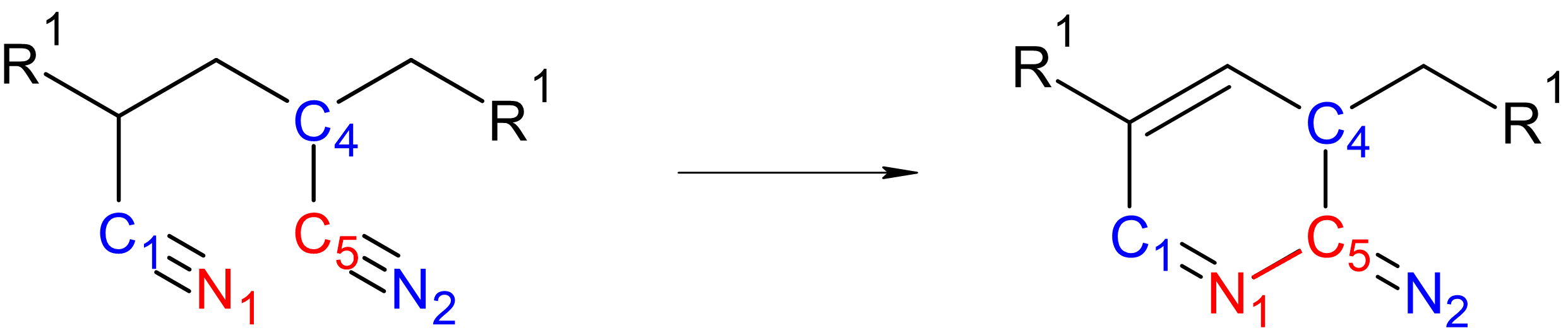}
        \caption{Model A}
        \label{fig:SystemA}
    \end{subfigure}
    \vspace{1em} 
    \begin{subfigure}[b]{\textwidth}
        \centering
        \includegraphics[scale=0.5]{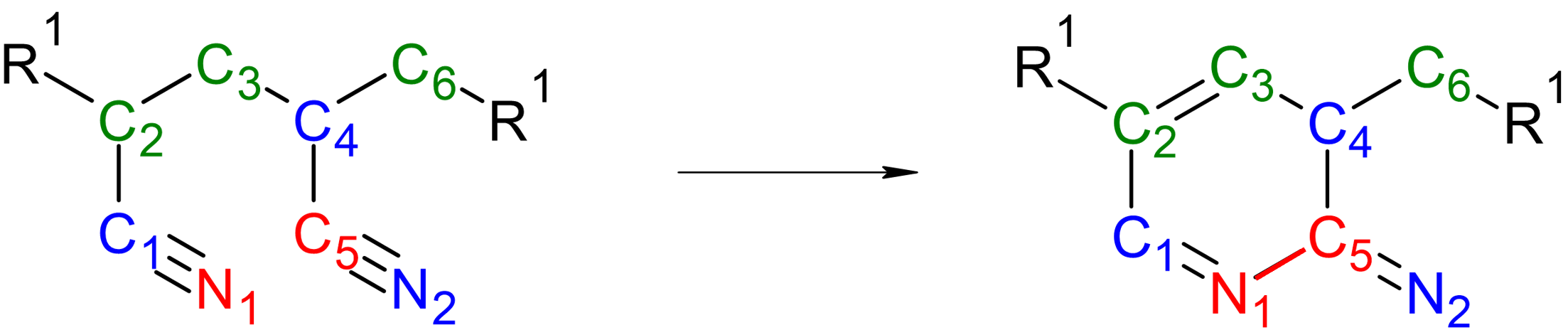}
        \caption{Model B}
        \label{fig:System B}
    \end{subfigure}
    \caption{In Model A (a), only the nearest neighbors (blue) of the reactive atoms (red) are considered, whereas in Model B (b), the analysis also includes the second-nearest neighbors (green).}
\end{figure}

We developed two models that span different numbers of atoms around the reactive site. As depicted in Fig.\ref{fig:SystemA}, Model A, includes the reactive atoms, C$_{5}$ and N$_{1}$ and the atoms directly bonded to them, C$_{1}$, C$_{4}$, N$_{2}$. Model B expands the scope to also include the second set of bonded atoms, C$_{2}$, C$_{3}$, and C$_{6}$. This addition increases the total number of atoms considered by the model from five to eight. Notably, Model B encompasses the entire cycle and an additional backbone carbon atom.

\subsection{Deep neural network and parameters}
To predict the strain energy associated with cyclization reactions, we employed a GNN as implemented in the open-source Allegro code \cite{musaelian_learning_2023}. Allegro is a local equivariant deep neural network designed to predict energies while preserving the physical symmetries of atomic systems. By enforcing invariance of the energy under translation, rotation, and reflection, it provides accurate and physically consistent predictions of atomic interactions. The model is trained to map the averaged pre-reaction positions of the considered atoms to the strain energy associated with the reaction. In this study, PAN cyclization is used as an illustrative example.
A total of 15,850 structure–energy pairs were used for each model, with 12,680 allocated for training and 1,585 frames for both validation and testing, resulting in a training/validation/test ratio of 80/10/10. To determine the optimal network architecture, six different models, varying in the number of hidden layers and features, were tested (see Table~\ref{tab:table_1}).
\begin{table}[h]
    \centering
    \begin{tabular}{c c c c c }
      Network Architecture & Hidden Layers & Features & Epochs & Learning Rate \\
      \hline
      I & \multirow{3}{*}{1} & 32 & \multirow{3}{*}{100} & \multirow{3}{*}{0.001}  \\
      II   &  & 64 &  &  \\
      III & & 128 & & \\  \hline
      IV  & \multirow{3}{*}{4}  & 32 & \multirow{3}{*}{100} & \multirow{3}{*}{0.001} \\ 
      V & &  64 & &  \\
      VI    &  &  128 &  &   \\ 
    \end{tabular}
      \caption{Network hyperparameters employed in the different architectures}
    \label{tab:table_1}
\end{table}
After training six models using different configurations—32, 64, and 128 features with either 1 or 4 hidden layers—the configuration with 128 features and a single hidden layer for Model A (system with 5 atoms) demonstrated the best predictive performance, effectively avoiding both underfitting and overfitting. For Model B (system with 8 atoms), the configuration with 128 features and a single hidden layer also showed reliable predictive accuracy. In contrast, the other architectures exhibited signs of underfitting or overfitting, showing discrepancies at high energies. Once trained, the energy is predicted using \textit{Nequip$_Calculator$} \cite{batzner_e3-equivariant_2022}, a tool developed by the same team behind Nequip.
\subsection{Reaction selection}
Following the approach developed by Yao et al. \cite{yao_molecular_2024}, reactions are dynamically identified as the system evolves. Model B is then used to predict reaction energies from pre-reaction coordinates. At each iteration, reactions are ranked according to their predicted energies, and only the 30\% with the lowest energies are retained. This adaptive filtering prioritizes energetically favorable pathways and effectively modulates the overall reaction rate.
\section{Results and discussion}
\label{Result}
\subsection{Effect of the local environment size on model accuracy} 
\begin{figure}[h]
    \begin{minipage}[c]{.46\linewidth}
        \centering \includegraphics[scale=0.52]{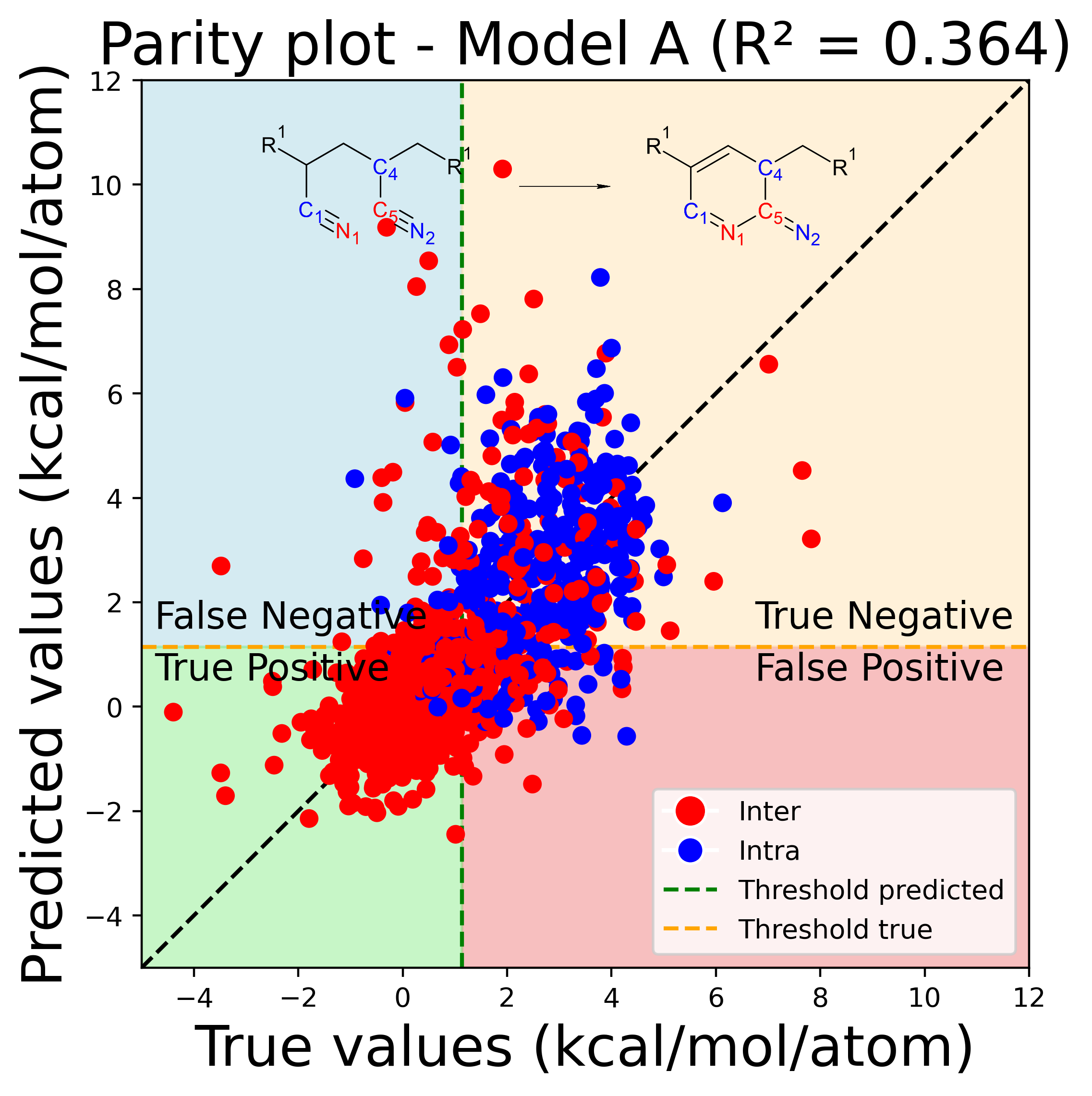}
        \vspace{2pt} 
        \text{(a)} 
    \end{minipage}
    \hspace{0.01cm}
    \begin{minipage}[c]{.46\linewidth}
        \centering
        \includegraphics[scale=0.5]{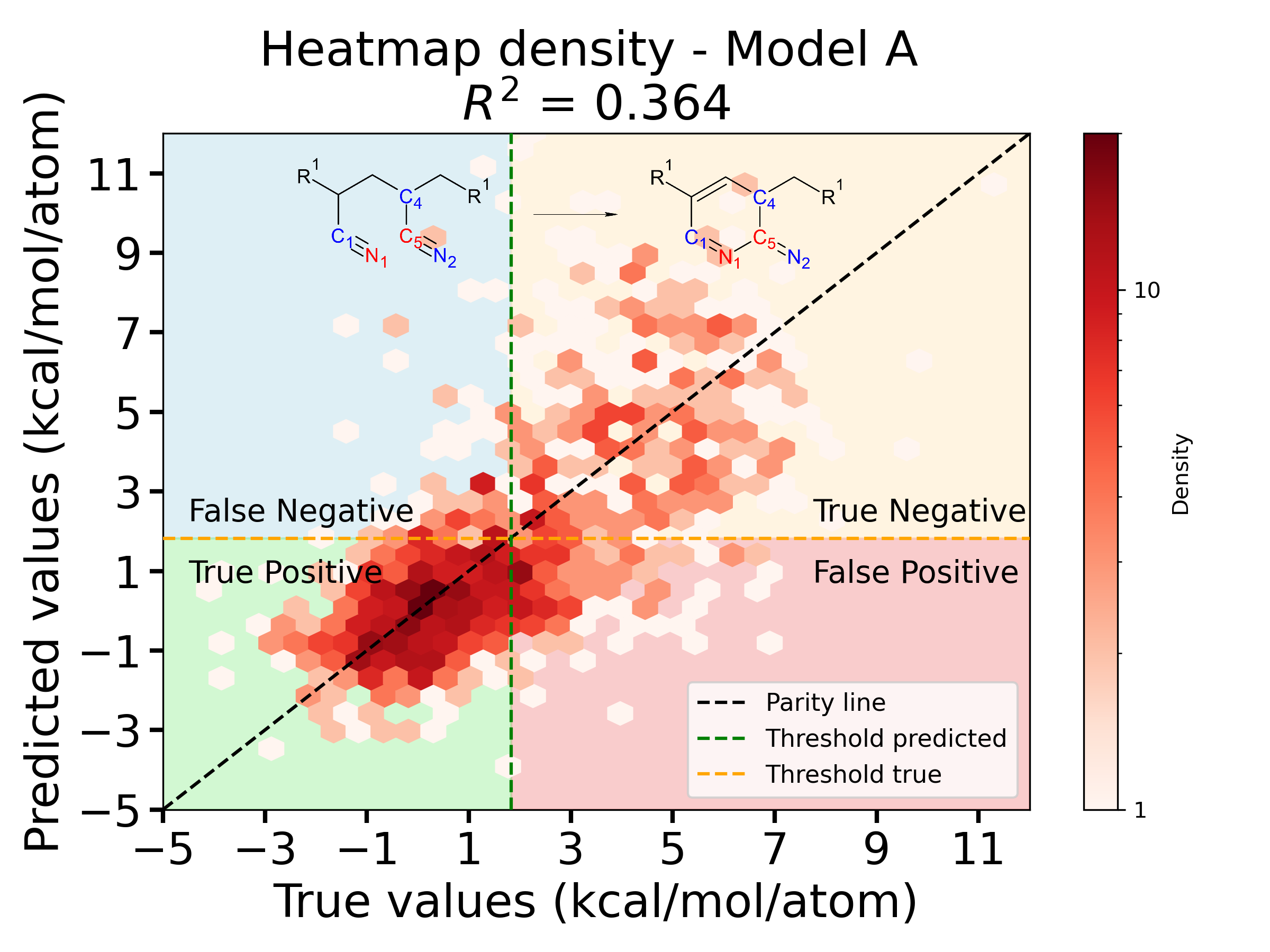}
        \vspace{2pt} 
        \text{(b)} 
    \end{minipage}
    \caption{(a) Parity plot of predicted and true values of strain energy from Model A. The four regions represent a classification problem used to assess the accuracy of the model described in Section 3.2. (b) Parity plot as a heat map representing the density distribution of the dataset.}
    \label{fig4:Heatmap_A}
\end{figure}

Figure~\ref{fig4:Heatmap_A}a, compares the strain energies predicted by Model A with the ground truth; only results from the test set are shown. Before discussing model performance, we note the broad distribution of energies sampled in the cyclization reactions, ranging from –2 to over 5 kcal/mol per atom. Similar trends have been reported in previous studies \cite{li_molecular_2010}, highlighting the importance of accounting for strain energy in chemical reaction rates in condensed systems. No significant difference is observed between intramolecular and intermolecular reactions.
The Heatmap in Figure \ref{fig4:Heatmap_A}b, reveals a cluster of reaction energies ranging from -3 to 3 kcal/mol/atom. The coefficient of determination r$^{2}$=0.364 indicates, indicates that Model A partially reproduces the overall relationship between predicted and true strain energies. While the model captures the general trend, particularly for low-energy configurations, a substantial portion of the variance remains unexplained. This suggests that although Model A identifies meaningful correlations and general trends in strain energy, it fails to accurately predict the strain energies of highly strained reactions.

\begin{figure}[H]
    \centering
    \includegraphics[scale=0.5]{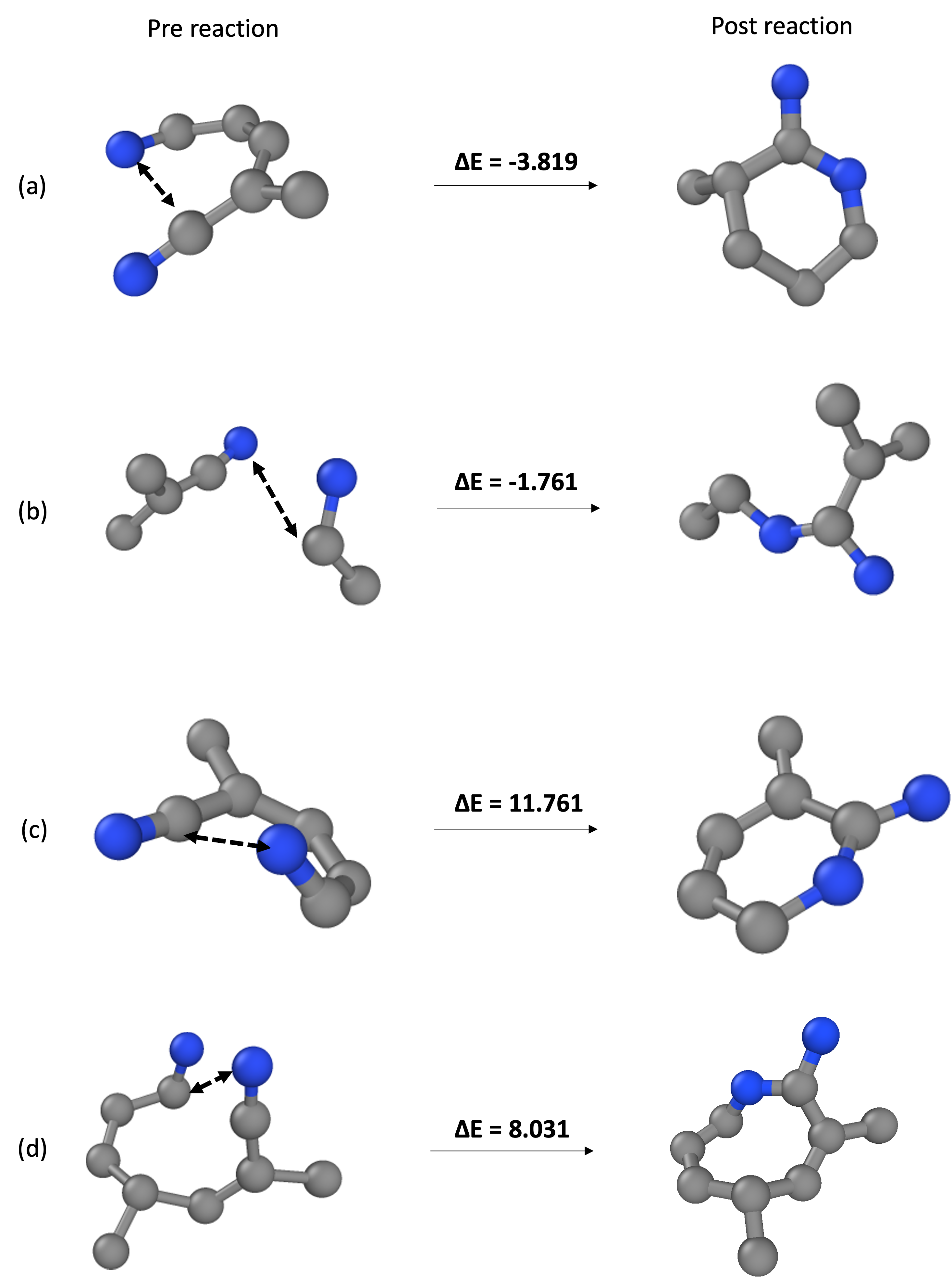}
    \caption{Subfigures (a) and (b) depict intramolecular and intermolecular pre-reaction configurations, respectively, that yield realistic reaction energies. In these cases, the reacting atoms are nearly coplanar, lying approximately within the same molecular plane, which minimizes pre-reaction strain. As a result, the corresponding geometries and reaction energies, $\Delta E$, are energetically favorable ($\Delta E < 0$). In contrast, subfigures (c) and (d) represent less favorable configurations, for which $\Delta E > 0$. Such configurations are energetically unfavorable, and the model is trained to identify and prevent them from reacting.}
    \label{fig:Bad_configurations}
\end{figure}

\noindent To motivate the introduction of Model B, we highlight representative low- and high-strain-energy configurations in Fig.~\ref{fig:Bad_configurations}. Panels (a) and (b) show examples of a cyclization reaction and an inter-chain reaction, respectively, both exhibiting low strain energies. Panels (c–d) illustrate high-strain-energy reactions. Reactions (d) and (e) produce topologies different from the desired cyclization, which motivated the choice of the geometrical criterion in Ref.~\cite{yao_molecular_2024} and explains their unfavorable energies. Interestingly, reaction (c) is a cyclization reaction with very high strain energy. Although the topology of reaction (c) is identical to that of (a), the ring is highly strained due to steric effects. Based on these observations, we speculated that including additional neighboring atoms could improve the model’s ability to distinguish between low- and high-energy configurations.

\begin{figure}[H]
    \begin{minipage}[c]{.46\linewidth}
        \centering \includegraphics[scale=0.52]{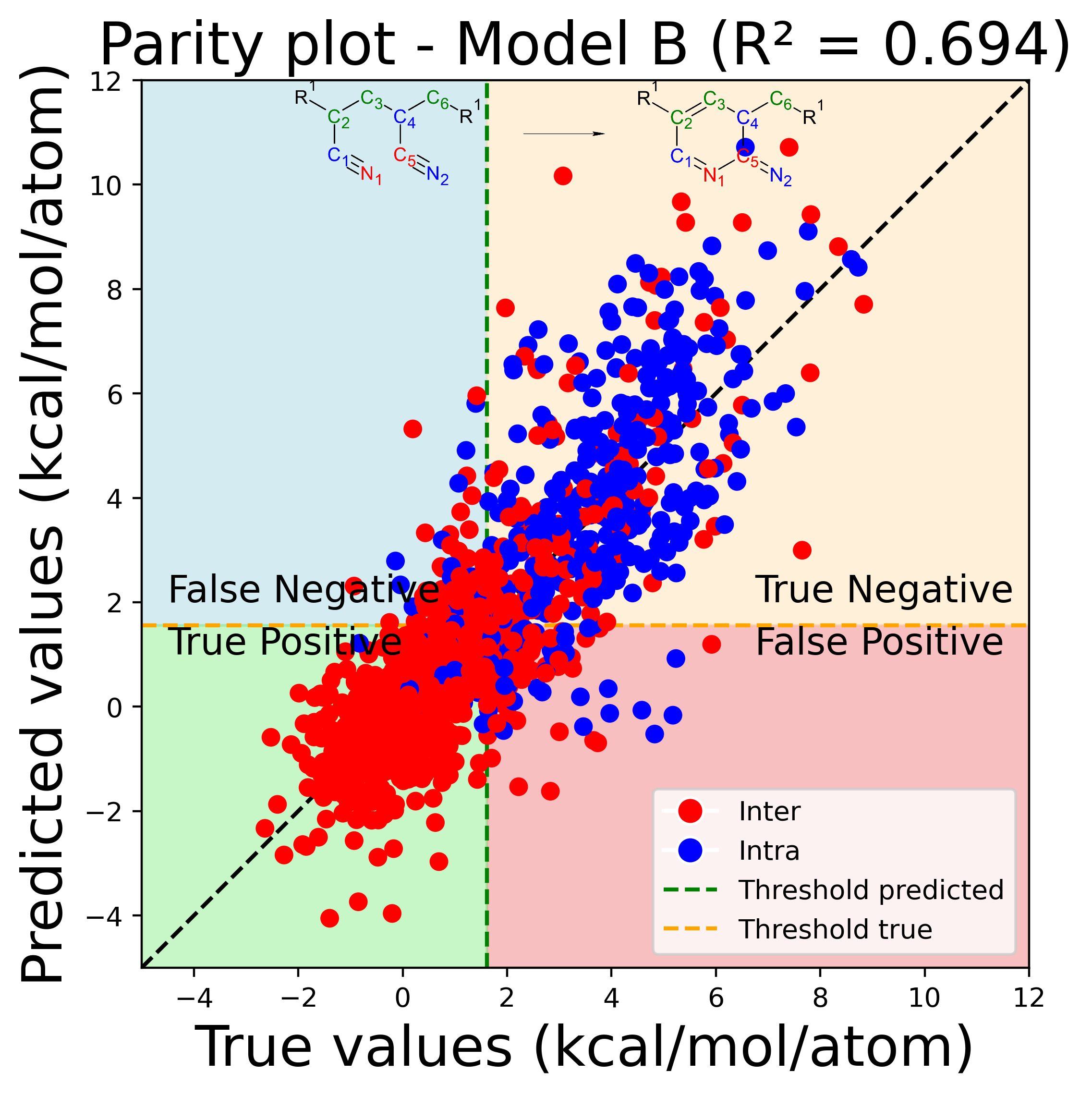}
        \vspace{2pt} 
        \text{(a)} 
        \label{fig6:parity_plotB}
    \end{minipage}
    \hspace{0.1 cm}
    \begin{minipage}[c]{.46\linewidth}
        \centering
        \includegraphics[scale=0.5]{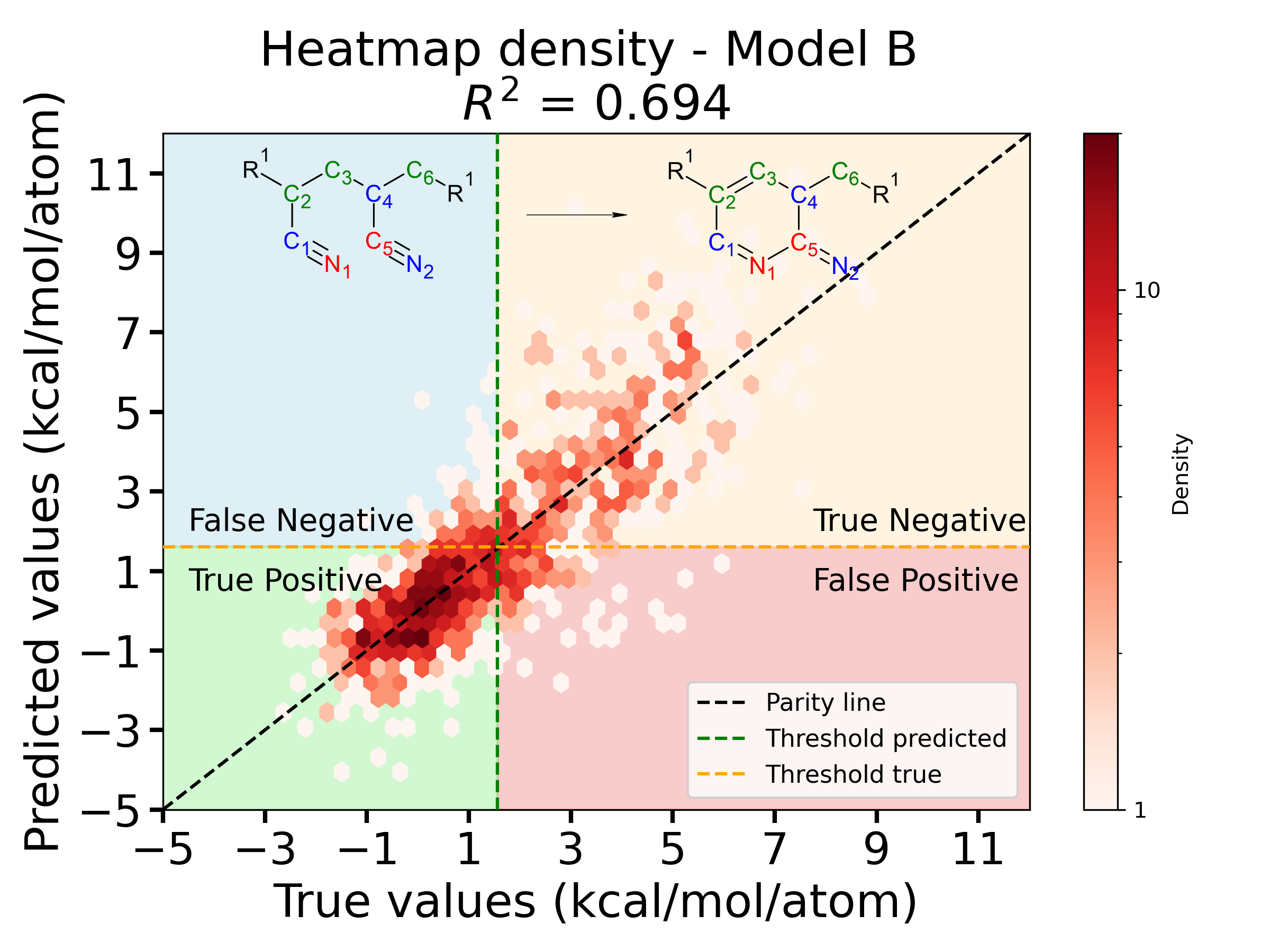}
        \vspace{2pt} 
        \textbf{(b)} 
    \end{minipage}
    \caption{The parity plot (a) shows improved alignment between predicted and true values, whereas the density heatmap (b) reveals a more diffuse distribution along the parity line, without forming distinct clusters.}
    \label{fig7:Heatmap_B}
\end{figure}
The parity plot of strain energies for Model B (Fig.~\ref{fig7:Heatmap_B}a) exhibits a markedly improved predictive performance compared to Model A (Fig.~\ref{fig4:Heatmap_A}a). In particular, Model B more accurately captures high-strain-energy reactions. This enhancement is attributed to the inclusion of two additional atoms from the polymer backbone, where the presence of sp$^{2}$-hybridized carbon atoms increases the chain stiffness and restricts the rotation of the C–CN groups, as discussed previously. The coefficient of determination (r$^{2}$ = 0.694) confirms a better agreement between predicted and true energies across both low- and high-energy regimes. Nevertheless, a residual cluster of reactions with strain energies between –2 and 3 kcal/mol/atom persists, though the energy distribution is now more symmetrically aligned along the parity line.
Table~2 summarizes the results for the two models. We note that the MAE per atom for the training and test sets is similar, indicating that the models neither overfit nor underfit.
\subsection{Model evaluation}
\begin{table}[h]
    \centering
    \begin{tabular}{c c c c c c c c}
    \hline
      Model & Atoms & Hidden & Features & Train  & Test  & bias & r$^{2}$  \\ 
        & & Layers &   &MAE$^{*}$ & MAE$^{*}$  & \\
    \hline
      A  & 5 & 1 & 128 & 1.651 & 1.640 & -0.15 & 0.364\\
      B & 8 & 1 & 128 & 1.01 & 1.02  & -0.03 & 0.694\\ 
    \hline
    \end{tabular}
    
    \footnotesize
    \textit{$^{*}$MAE are normalized per atom (5 in model A and 8 in model B and expressed in kcal/mol/atom})
    \caption{Parameters and performance comparison of the two models. }
\label{tab:Parameter}
\end{table}
\noindent
Comparing the performance of model A and model B shows a clear improvement in the predictions of the latter, with further details provided in the Supplementary Information. For model A, the MAE is 1.651 kcal/mol/atom, with a slightly negative biases (-0.15) and an r² of only 0.364, indicating that the model explains very little of the variance in the data. In contrast, model B displays an MAE of 1.01 kcal/mol/atom, a near-zero biases (-0.03), and an r² of 0.694, showing better overall accuracy, lower error dispersion, and a significant ability to capture data trends.
These results suggest that the adjustments made for model B have significantly improved the reliability and consistency of the model's predictions. The purpose of this study is to develop models capable of adjusting reaction rates for condensed-phase systems by using utilize the Bell–Evans–Polanyi relationship.\cite{evans_inertia_1938, evans_applications_1935} between activation energy and the overall energy of chemical reactions to adjust relative reaction rates. To assess the ability of the models to limit reactions that would result in high strain energies, we reformulate the problem as a classification task, aiming to identify high-energy reactions to be avoided. A strain energy cutoff of 2 kcal/mol per atom is used (see shaded regions in Figs.~\ref{fig7:Heatmap_B}).
True positives (TP) occur when the model correctly predicts an energy below 2 kcal/mol/atom, while true negatives (TN) correspond to correctly predicted energies above the threshold. False positives (FP) and false negatives (FN) are defined similarly. Classification performance is evaluated using standard metrics: sensitivity (recall) as TP / (TP + FN), specificity as TN / (TN + FP), precision as TP / (TP + FP), and accuracy as (TP + TN) / (TP + TN + FP + FN). The results are summarized in Table~\ref{tab:Result}.
Model B demonstrates a better balance across key metrics. Sensitivity, which measures the proportion of actual positives correctly identified, is slightly lower for Model A (82.12\%) compared to Model B (85.80\%). However, Model B outperforms Model A in specificity (85.82\% vs. 78.62\%), indicating a stronger ability to correctly identify negative cases—crucial for minimizing false positives in reaction site detection. Precision, reflecting the proportion of predicted positives that are correct, is also higher for Model B (79.90\% vs. 70.06\%). Furthermore, the false positive rate is significantly reduced in Model B (14.18\% vs. 21.31\%), and overall accuracy improves (85.8\% vs. 74.98\%). In addition to that Model B exhibits a slightly lower false negative rate (14.20\% vs. 17.89\%), its superior precision and specificity make it more suitable for applications where false positives must be minimized.
\begin{table}[h]
    \centering
    \begin{tabular}{ccc}
     \hline
        Parameters   & Model A  & Model B  \\ \hline
        Sensitivity  & 82.12 & 85.80 \\
        Specificity  & 78.62 & 85.82 \\
        Precision    & 70.06 & 79.90 \\
        False negative rate & 17.86 & 14.20 \\ 
        False positive rate & 21.31 & 14.18\\
        Accuracy     & 74.98 & 85.8 \\ \hline
        \end{tabular}
    \caption{Comparison of performance metrics between Model A and Model B.}
    \label{tab:Result}
\end{table}
\subsection{Reaction rate adjustment}
In the proposed framework, only the pre-reaction atomic coordinates are required as input. The trained GNN directly predicts whether a reaction is likely to occur, thus avoiding the computationally expensive post-processing steps usually involved in nonreactive force-field simulations.
From a computational perspective, the system explored in this study involves more than 15,000 potential reactions, achieving approximately 90 \% conversion with an average of 10 reactions per cycle (about 10\% of all possible events). Performing explicit potential energy calculations between pre- and post-reaction configurations, as implemented in algorithms such as REACTER\cite{gissinger_molecular_2024}, would increase the total computational cost by nearly an order of magnitude. Consequently, the machine learning approach provides a scalable and efficient alternative for large molecular systems with numerous reactive sites. 
\begin{figure}[H]
    \centering
    \begin{minipage}[t]{0.46\linewidth}
        \centering
        \includegraphics[scale=0.489]{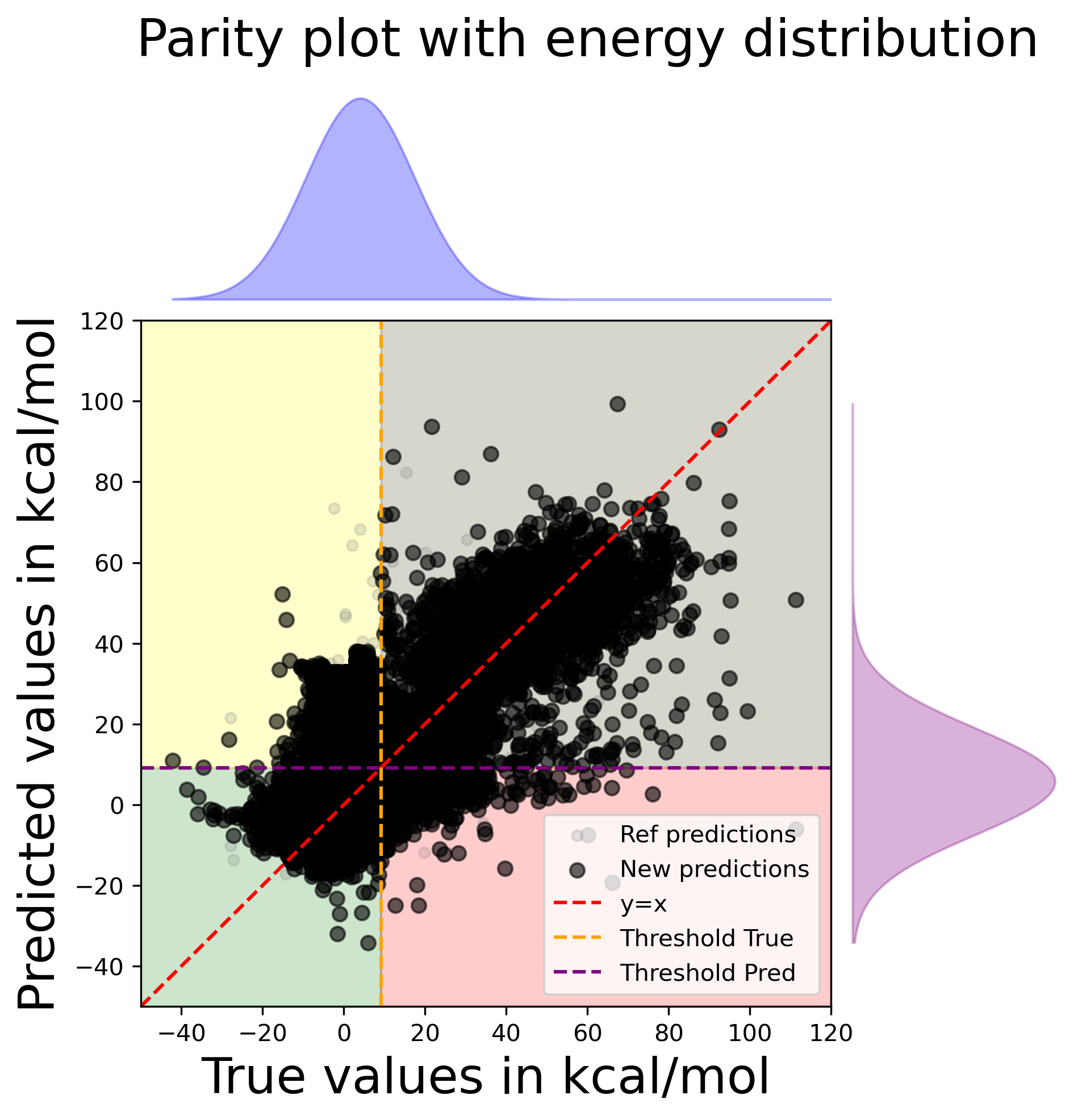}
        \label{fig:10}
        \vspace{2pt} 
        \text{(a)} 
    \end{minipage}
    \hspace{0.05\linewidth} 
    \begin{minipage}[t]{0.46\linewidth}
        \centering
        \includegraphics[scale=0.484]{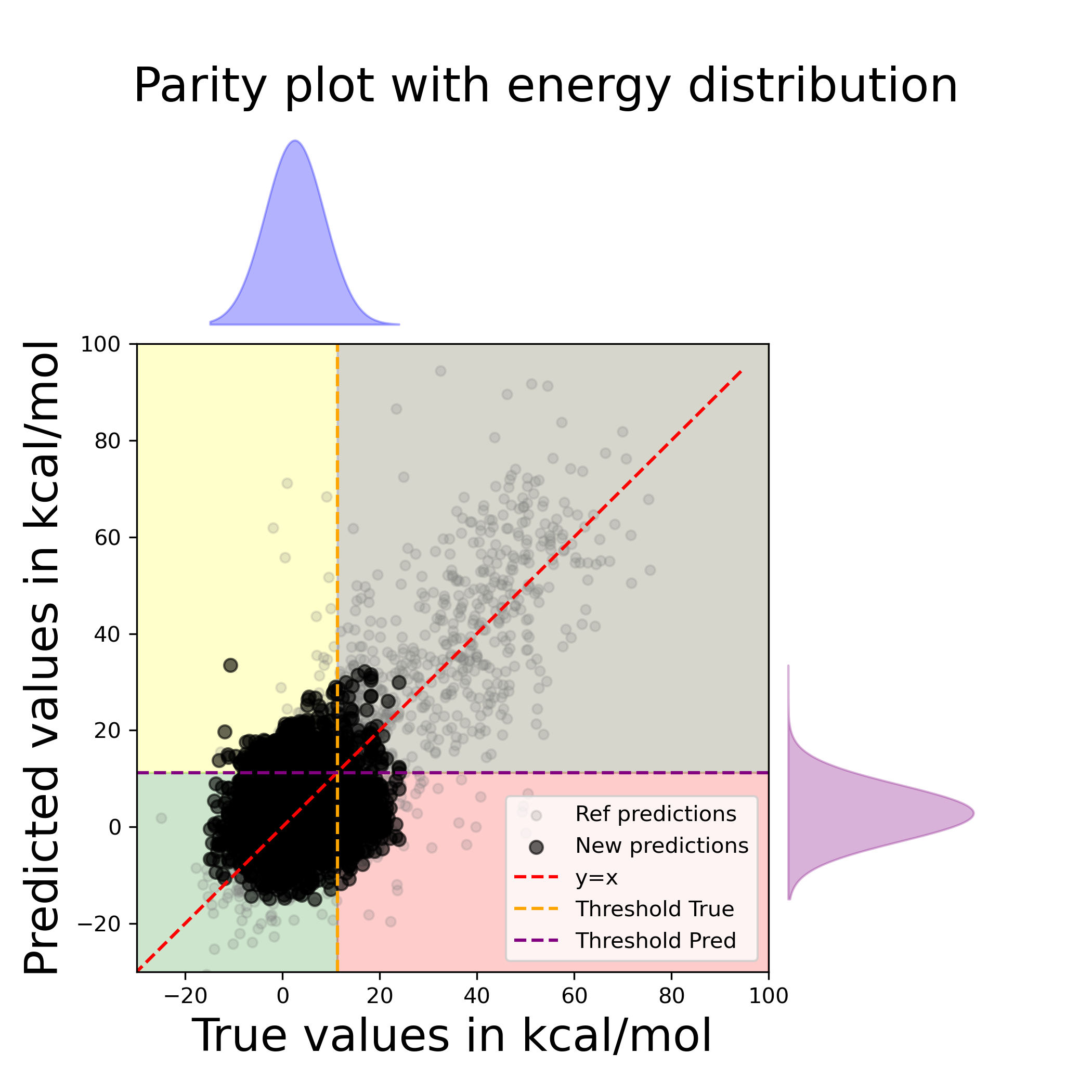}
        \vspace{2pt} 
        \text{(b)} 
        \label{fig:11}
    \end{minipage}
    \caption{Comparison of PAN polymerization with and without an energy-based selection criterion. The parity plot (a), without any energy filtering, shows a wide and broad distribution of reaction energies. In contrast, the parity plot (b), with the energy criterion applied, demonstrates that the selection favors the lowest-energy configurations, leading to a more concentrated distribution.}
\label{fig:parity_comparison}
\end{figure}
\noindent
To identify the most relevant reactions, we applied Model B, which was previously trained, to perform an on-the-fly selection based on predicted strain energies (fig \ref{fig:parity_comparison}a -\ref{fig:parity_comparison} b). Specifically, we retained the 30\% of reactions with the lowest predicted energies as mentioned in section Methods. This energy-based criterion ensures that the selected subset emphasizes low-energy, physically relevant configurations, allowing subsequent analyses to focus on reactions that are most critical for understanding reaction behavior. We observe that when no energy-based selection criterion is applied, i.e., when reactions are chosen purely based on geometry for the same conversion fraction (here 30\%), the distribution of true strain energies is very broad, ranging from -40 to 100 kcal/mol/atom. In contrast, applying an energy-based selection results in a much narrower distribution, tightly centered around -30 kcal/mol/atom. This demonstrates that energy-based filtering effectively focuses on low-energy, physically relevant reactions, reducing the variance in the selected dataset. 

\section{Conclusions}

State of the art molecular-level simulations of reactive processes in condensed phases use a geometrical criteria to select possible reactions and MD to relax the structures following reaction cycles. An analysis of prior indicate a wide range of local strain energies associated with the chemical steps which could result in unrealistic structures. We collected data of 15,000 reactions from our group's prior work and found that a graph neural network can accurately predict the strain energy associated with the reaction using the pre-reaction coordinates as only input. The model is computationally efficient and can be incorporated in molecular simulations to provide accurate estimate of strain energies and correct the reaction rates of possible reaction. Avoiding chemical reactions that would result in high strains would lead to more realistic structures in simulations of polymerization, depolymerization, and other processing steps in molecular materials.

\section{Data availability}
The source code of PAN stabilization simulation along with an example run is available at
also a code to extract the energy is available at https://github.itap.purdue.edu/StrachanGroup/
CondPhaseChemModeler.
\section{Supporting information}
Model hyperparameters, training choices, and performance evaluation.
\section{Acknowledgments}
This work was supported by Syensqo specialty polymer. Computational resources of Purdue University and nanoHUB are gratefully acknowledged
\printbibliography
\end{document}